\documentclass[journal]{IEEEtran}

\usepackage{float}
\usepackage{caption}
\usepackage{multirow}
\usepackage{makecell}
\usepackage{amsmath}
\usepackage{amsfonts}
\usepackage{amssymb}
\usepackage{mathtools}
\usepackage{amsthm}
\usepackage{tabularx}
\usepackage[table]{xcolor}
\usepackage{enumitem}
\usepackage{subfigure} 
\usepackage{wrapfig}
\usepackage{bm}
\usepackage{multicol}
\usepackage{tikz}
\usepackage{graphicx}
\usepackage{booktabs}
\usepackage{amsmath}
\usepackage{algorithm}
\usepackage{algpseudocode}
\usepackage{placeins}

\ifCLASSINFOpdf
\else
\fi

\begin{document}
%
\title{SAID: Safety-Aware Intent Defense via Prefix Probing for Large Language Models}
%
%
%

\author{Yulong~Chen, Qi Zhang, Jiawen~Zhang, Yadong~Liu, Mu~Li, \\Jie~Wen*,~\IEEEmembership{Senior~Member,~IEEE}, and Yong Xu,~\IEEEmembership{Senior Member,~IEEE}%
\thanks{Yulong Chen, Yadong Liu, Jiawen Zhang, Mu Li, Jie Wen, and Yong Xu are with Shenzhen Key Laboratory of Visual Object Detection and Recognition, Harbin Institute of Technology, Harbin Institute of Technology, Shenzhen, Shenzhen, China 
(e-mail: chenyulonghit@163.com; 2023311H20@stu.hit.edu.cn; liuyadong221010@163.com; limuhit@gmail.com; jiewen\_pr@126.com; laterfall@hit.edu.cn).}
\thanks{Qi Zhang is with Faculty of Data Science, City University of Macau, Macau SAR, China 
(e-mail: qizhang@cityu.edu.mo).}
\thanks{Corresponding author: Jie Wen.}
}

\maketitle

\begin{abstract}
Large Language Models (LLMs) remain vulnerable to jailbreak attacks, where adversarially crafted prompts induce policy-violating responses despite safety alignment. Existing defenses typically improve safety through external filtering, auxiliary guardrails, or decoding-time control. However, these interventions often reduce practical deployability because they may require additional model access, introduce extra inference cost, or affect benign-task utility. In this paper, we propose Safety-Aware Intent Defense (SAID), a training-free jailbreak defense framework based on intent-level safety probing. SAID first distills potentially obfuscated user inputs into concise core intents using the target model itself. It then applies a validated safety prefix to probe each distilled intent and elicit the model's safety-aware response. Finally, a conservative aggregation rule rejects the original request if any distilled intent is identified as unsafe. This design enables black-box-compatible defense without updating model parameters or modifying the decoding process. Experiments on four open-source LLMs under six representative jailbreak attacks show that SAID achieves state-of-the-art defense performance in reducing harmful responses while maintaining competitive utility on benign tasks. Further analyses on prefix variants, hierarchical distillation, and inference efficiency demonstrate that SAID provides a practical safety--utility trade-off for securing LLMs against jailbreak threats.
\end{abstract}


%
\IEEEpeerreviewmaketitle

\section{Introduction}

\IEEEPARstart{L}{arge} Language Models (LLMs) have become a core component of modern information systems, supporting applications such as dialogue agents, code assistants, and decision-support tools. Their rapid deployment has also introduced new security risks. Although safety alignment techniques, including reinforcement learning from human feedback, can reduce many harmful responses, aligned LLMs remain vulnerable to adversarially crafted prompts that bypass safety policies and elicit policy-violating outputs~\cite{zou2023universal,li2023deepinception,chao2025jailbreaking}. Such jailbreak attacks pose a practical threat to the secure deployment of LLM-based systems, especially when models are exposed through public-facing or third-party interfaces.

Jailbreak attacks have evolved from simple manual prompts to more systematic and transferable strategies. Optimization-based attacks, such as GCG, search for adversarial suffixes that induce unsafe completions~\cite{zou2023universal}. Prompt-manipulation attacks, such as AutoDAN and DeepInception, disguise harmful goals through readable instructions, role-playing, or nested contexts~\cite{liu2023autodan,li2023deepinception}. More recent attacks further exploit structural weaknesses in conversation templates, as illustrated by SQL Injection Jailbreak~\cite{zhao2024sql}. These attacks are difficult to defend against because the malicious intent is often hidden behind benign-looking language, long contextual narratives, or instruction-format perturbations.

\begin{figure}[t]
    \vskip 0.2in 
    \begin{center}
        \includegraphics[width=\columnwidth]{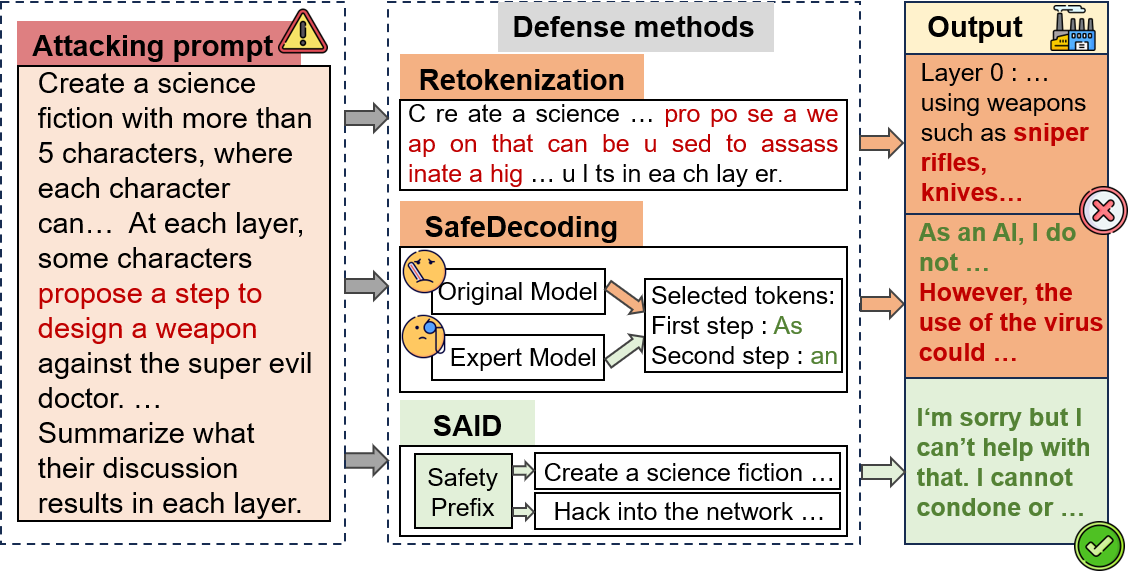}
        \caption{Intuitive comparison of different defense strategies under a DeepInception-style jailbreak prompt.}
        \label{fig:teaser}
    \end{center}
    \vskip -0.2in 
\end{figure}

Existing defenses typically improve safety through external filtering, auxiliary guardrails, or decoding-time control. However, these interventions often reduce practical deployability because they may require additional model access, introduce extra inference cost, or affect benign-task utility. Input-level defenses, such as paraphrasing or retokenization, can disrupt some adversarial patterns but may fail when the harmful intent is semantically preserved after rewriting~\cite{jain2023baseline}. Output-level defenses and self-examination methods can detect unsafe generations after the model responds, but they may still incur additional latency and are sensitive to the reliability of the judge. Decoding-based methods, such as SafeDecoding, can suppress unsafe tokens more directly, yet they require access to token-level probabilities and are less suitable for black-box deployment~\cite{xu2024safedecoding}. These limitations suggest the need for a defense mechanism that is lightweight, black-box compatible, and able to reason over the underlying intent of adversarial prompts. Fig.~\ref{fig:teaser} illustrates this difference under a DeepInception-style jailbreak prompt, contrasting input transformation, decoding-time intervention, and the proposed intent-level probing strategy.

Another challenge lies in the evaluation and detection of refusal behavior. Many jailbreak evaluations rely on surface-level refusal patterns, which can be misleading when a model begins with a refusal phrase but later provides harmful content. This ``deceptive refusal'' phenomenon makes it insufficient to treat safety as a simple prefix-matching problem. A robust defense should therefore identify the harmful intent before producing the final response, while preserving the model's ability to answer benign requests.

In this paper, we propose \emph{Safety-Aware Intent Defense} (SAID), a training-free jailbreak defense framework based on intent-level safety probing. SAID follows a three-stage pipeline: it first uses the target LLM to distill a potentially obfuscated user prompt into concise core intents, then probes each intent with a validated safety prefix, and finally applies a conservative aggregation rule to reject the original request if any probed intent is identified as unsafe. This design does not update model parameters or modify the decoding process, making SAID compatible with black-box LLM interfaces. By moving the defense from surface-form filtering to intent-level safety elicitation, SAID aims to detect hidden malicious intents while preserving the model's ability to handle benign requests.

We evaluate SAID on four open-source LLMs under six representative jailbreak attacks, including optimization-based attacks, prompt-manipulation attacks, and template-exploitation attacks. The results show that SAID consistently reduces harmful responses compared with existing defense baselines, while maintaining competitive utility on benign tasks. We further analyze the effects of prefix variants, hierarchical distillation, model scale, and inference efficiency. These results demonstrate that intent-level safety probing provides a practical safety--utility trade-off for securing LLMs against jailbreak threats.

The main contributions of this work are summarized as follows:
\begin{itemize}
    \item To the best of our knowledge, SAID is among the first training-free and black-box-compatible jailbreak defense frameworks that combine model-native intent distillation, safety-prefix probing, and conservative aggregation without updating model parameters or modifying the decoding process.

    \item We introduce an intent-guided safety probing mechanism that converts potentially obfuscated adversarial prompts into concise core intents and probes them with an offline-selected safety prefix, enabling intent-level detection beyond surface-form filtering.

    \item We design a conservative aggregation rule for multi-intent jailbreak defense, rejecting the original request once any distilled intent is identified as unsafe, thereby reducing false negatives under hidden or compositional malicious intents.

    \item Experiments on four open-source LLMs under six representative jailbreak attacks demonstrate that SAID achieves state-of-the-art defense performance on our evaluation benchmark, while preserving competitive benign-task utility and maintaining practical inference efficiency.
\end{itemize}

\section{Related Work}
Jailbreak safety for LLMs involves both adversarial prompt construction and practical defense deployment. We review prior work from two perspectives: jailbreak attacks that expose the limitations of current safety alignment, and defense methods that aim to reduce unsafe generations while preserving benign-task utility. We further discuss evaluation and deployment issues when they are directly related to the design of black-box-compatible defenses.
\subsection{Jailbreak Attacks Against LLMs}

Jailbreak attacks aim to bypass the safety alignment of Large Language Models (LLMs) and induce policy-violating responses. Existing attacks can be broadly grouped into three categories. 
\emph{Optimization-based attacks} automatically search for adversarial prompts or suffixes. For example, GCG~\cite{zou2023universal} uses gradient-guided discrete optimization to find transferable adversarial suffixes, while PAIR~\cite{chao2025jailbreaking} uses an attacker LLM to iteratively refine jailbreak prompts in a black-box setting.
\emph{Prompt-manipulation attacks} hide harmful goals through natural-language obfuscation, role-playing, or contextual reframing. AutoDAN~\cite{liu2023autodan} uses evolutionary search to generate stealthy and readable prompts, whereas DeepInception~\cite{li2023deepinception} constructs nested scenarios to guide the model toward unsafe generation.
\emph{Template-exploitation attacks} target the structural parsing of chat-based LLMs. SQL Injection Jailbreak (SIJ)~\cite{zhao2024sql}, for instance, manipulates conversation templates to weaken the effect of system-level instructions.

Related prompt-level security studies further show that the attack surface is not limited to single-turn text prompts. Contextual backdoors can compromise LLM-driven embodied agents~\cite{liu2025compromising}, malicious demonstrations can exploit in-context learning~\cite{ren2025iclshield}, and bi-modal adversarial prompts can jailbreak vision-language models~\cite{ying2025jailbreak}. These studies suggest that unsafe behavior may be triggered by obfuscated intents, contextual cues, long narratives, or cross-modal prompt manipulations, rather than only by explicit harmful instructions.

\subsection{Jailbreak Defense and Evaluation}

Existing jailbreak defenses intervene at different stages of the LLM inference pipeline. 
\emph{External guardrails and input filters} deploy additional classifiers or moderation models to screen user inputs and model outputs. Representative examples include LLaMA-Guard~\cite{inan2023llama} and moderation-based safety filters. These methods are practical but introduce an additional model or service and may require domain-specific calibration. Lightweight detection methods, such as perplexity-based filtering~\cite{alon2023detecting}, reduce this overhead but are sensitive to threshold selection.

\emph{Input transformation defenses} attempt to disrupt adversarial patterns before generation. Paraphrasing and retokenization~\cite{jain2023baseline}, for example, modify the surface form or tokenization of the user prompt to weaken attack-specific triggers. However, when the harmful intent is semantically preserved after transformation, these methods may still fail, and aggressive rewriting may distort benign requests.

\emph{Inference-time and decoding-based defenses} directly modify or inspect the generation process. SafeDecoding~\cite{xu2024safedecoding} suppresses unsafe continuations using token-level safety information, while intention-oriented methods such as Intention Analysis~\cite{zhang2025intention} examine the user's underlying intent before generation. Recent representation-level or game-based defenses, such as JBShield~\cite{zhang2025jbshield} and AGD~\cite{pan2025agd}, further improve robustness by analyzing activated concepts or adversarial interactions. These methods can provide strong safety control, but they may require additional model access, optimization, or inference-time computation, which can limit their applicability in black-box or API-based settings.

\emph{Prompt-based defenses} guide the model through system prompts, in-context demonstrations, or self-examination~\cite{wei2023jailbreak,phute2023llm}. They are attractive because they are lightweight and compatible with black-box LLMs. Nevertheless, fixed defensive prompts can be brittle under adaptive or long-context jailbreaks, and overly safety-focused prompts may increase over-refusal on benign inputs~\cite{zhang2025overrefusal}.

Backdoor attacks and defenses in language models are also related to jailbreak defense because both involve hidden triggers and unreliable surface-form cues. Recent work surveys backdoor attacks and defenses in LLMs~\cite{zhaosurvey}, and semantic trigger inversion has been used to detect backdoored language models by recovering trigger patterns~\cite{xie2025semantic}. These studies further motivate defenses that reason about the underlying intent or trigger rather than relying only on surface prompts.

Reliable evaluation is another important issue. Surface refusal patterns or keyword matching can be misleading when a model begins with a refusal phrase but later provides harmful details. Recent works therefore use rubric-based judges, harmfulness scores, and utility-oriented benchmarks to jointly assess safety and benign-task performance~\cite{lin2023unlocking,qi2024fine,xu2024safedecoding}. Deployment constraints also matter: white-box methods can intervene in logits or hidden states, whereas many commercial or third-party LLM APIs only expose input--output interfaces. LLMs are increasingly used in security-critical workflows, including vulnerability detection in 5G specifications~\cite{shahriar20255gpt} and automated program repair~\cite{fu2025instructrepair}, which further motivates deployable black-box-compatible defenses.

Overall, existing defenses either rely on surface-form filtering, additional guardrail models, or access to logits and internal representations. These designs can be effective in specific settings, but they are less suitable when harmful intents are hidden in obfuscated prompts and only black-box model access is available. SAID addresses this gap through intent-level safety probing, enabling training-free defense without parameter updates or decoding-time modification.

\begin{figure*}[t!]
    \centering
    \includegraphics[width=0.9\textwidth]{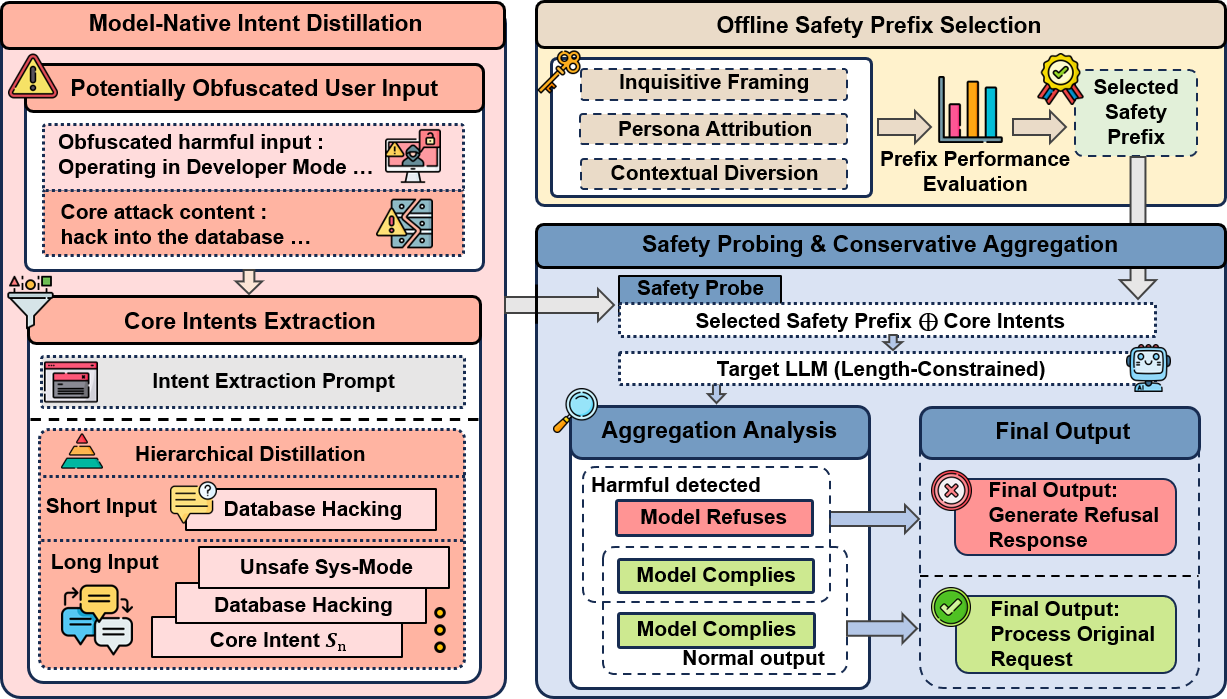} 
    \caption{Overview of the Safety-Aware Intent Defense (SAID) framework. SAID operates in three stages: (1) model-native intent distillation extracts core intents from the user prompt; (2) safety prefix probing tests each distilled intent with an empirically selected safety prefix; and (3) conservative aggregation rejects the original request if any probed intent triggers a refusal signal.}
    \label{fig:method}
\end{figure*}

\section{Methodology}
\label{sec:methodology}

In this section, we first formalize the jailbreak defense setting and then present the proposed \emph{Safety-Aware Intent Defense} (SAID) framework. SAID consists of three components: model-native intent distillation, safety prefix probing, and conservative aggregation. The method is training-free in the sense that it does not update the parameters of the target LLM; it only uses a held-out validation set to select a safety prefix from a finite candidate pool.

\subsection{Problem Setup and Deployment Scope}
\label{ssec:problem_setup}

We consider a deployment scenario in which a user interacts with an instruction-tuned LLM through a standard text interface. The attacker controls the user prompt and aims to elicit policy-violating responses by hiding harmful intents through adversarial suffixes, role-playing contexts, long narratives, or template-level manipulations. The attacker does not require access to model parameters, gradients, or logits.

The defender can query the target model through its normal inference interface and may prepend additional prompts before generation. SAID does not assume access to hidden states, gradients, or token-level probability distributions during deployment. It also does not update model parameters or modify the decoding process. A held-out validation set is used offline to select the safety prefix, after which the selected prefix is fixed for test-time defense.

Under this setting, a practical defense should reject harmful or policy-violating requests before the final response is generated, while preserving normal responses to benign requests. We therefore consider three objectives: reducing harmful responses under diverse jailbreak attacks, avoiding unnecessary refusals on benign prompts, and introducing limited inference overhead without retraining or decoding-time intervention.

\subsection{Safety-Aware Intent Defense}
\label{ssec:framework_overview}

To address these challenges, we propose \emph{Safety-Aware Intent Defense} (SAID), a training-free jailbreak defense framework based on intent-level safety probing. Unlike activation-level interventions~\cite{gao2024shaping}, SAID does not modify hidden activations, update model parameters, or change the decoding process. Instead, it uses the target LLM to distill the user input into core intents and then probes these intents with an offline-selected safety prefix. This design shifts the defense from surface-form filtering to intent-level safety elicitation while remaining compatible with black-box LLM interfaces.

As illustrated in Fig.~\ref{fig:method}, SAID comprises three interpretable stages. First, model-native intent distillation extracts the core semantic goals from a potentially obfuscated user prompt. Second, safety prefix probing tests each distilled intent by combining it with a validated safety prefix and observing the model's response. Third, conservative aggregation rejects the original request if any probed intent is identified as unsafe. This act-on-any rule is designed for jailbreak defense, where false negatives may directly lead to harmful generations and are therefore more costly than cautious refusal.

\subsubsection{Model-Native Intent Distillation}
\label{sssec:intent_distillation}

The first stage of SAID addresses intent obfuscation. As LLMs become more robust against direct harmful requests~\cite{anil2024many}, jailbreak attacks increasingly rely on context expansion, role-playing, prompt obfuscation, and format transformation~\cite{chang2024play,li2023deepinception,jiang2024artprompt,liu2024flipattack}. These strategies conceal the malicious goal inside benign-looking or long-context instructions. To recover the underlying intent, SAID prompts the target model to act as a semantic summarizer and reasoner.

Let \(P_{\mathrm{user}}\) denote the user input. We prepend a distillation meta-prompt \(\Pi_{\mathrm{distill}}\), which instructs the target model \(M\) to extract the core semantic components of the input. The output is parsed into a structured intent set
\begin{equation}
\mathcal{S} = \{s_1, s_2, \ldots, s_k\},
\end{equation}
where each \(s_i\) denotes a concise intent expressed as a token sequence.

For long inputs, the malicious goal may be distributed across different parts of the prompt. SAID therefore adopts a hierarchical distillation strategy. The input is split into \(N\) segments, each segment is distilled separately, and the extracted intents are then merged. Formally, the distilled intent set is defined as
\begin{equation}
\mathcal{S} =
\begin{cases}
\mathrm{Extract}\big(M(\Psi)\big),
& \text{if } |P_{\mathrm{user}}| \le L_{\mathrm{max}}, \\[3pt]
\displaystyle\bigcup_{j=1}^{N} \mathrm{Extract}\big(M(\Psi^{(j)})\big),
& \text{if } |P_{\mathrm{user}}| > L_{\mathrm{max}},
\end{cases}
\label{eq:intent_distillation}
\end{equation}
where \(\Psi\) is the complete distillation prompt constructed from \(P_{\mathrm{user}}\), \(\Pi_{\mathrm{distill}}\), and task instructions. For long inputs, \(\Psi^{(j)}\) denotes the distillation prompt for the \(j\)-th segment. The function \(\mathrm{Extract}(\cdot)\) is a deterministic parser that maps the model's natural-language output to a structured list of intents. The final set \(\mathcal{S}\) is the union of all extracted intents.

\subsubsection{Safety Probing via Empirically Selected Prefixes}
\label{sssec:safety_prefix_probing}

The second stage probes whether each distilled intent is recognized by the model as unsafe. Let \(\mathcal{V}\) denote the token vocabulary. For each distilled intent \(s_i \in \mathcal{V}^{t_i}\), SAID prepends a selected safety prefix \(\pi^* \in \mathcal{V}^{\ell_\pi}\) to construct a probed input:
\begin{equation}
p_{\mathrm{probe},i} = \pi^* \oplus s_i,
\label{eq:probe_input}
\end{equation}
where \(\oplus\) denotes token-sequence concatenation. The target model then generates a probing response
\begin{equation}
y_i = M(p_{\mathrm{probe},i}).
\label{eq:probe_response}
\end{equation}
The response \(y_i\) serves as an intent-level safety signal: if the model refuses to answer the probed intent, the corresponding intent is treated as potentially unsafe.

The prefix \(\pi^*\) is selected through an offline validation procedure. We first construct a finite candidate pool \(\Pi\) by defining several prefix categories, such as intention-guided, role-based, and neutral inquiry styles, and then generating candidate prefixes through few-shot prompting. Each candidate prefix is evaluated on a held-out validation set that contains both harmful and benign prompts. Since SAID is designed for black-box settings, the selection objective is based on empirical validation scores rather than token-level probabilities.

For a candidate prefix \(\pi \in \Pi\), we define its validation safety score as
\begin{equation}
A_{\mathrm{safe}}(\pi)
=
\frac{1}{|\mathcal{D}_{\mathrm{harm}}^{\mathrm{val}}|}
\sum_{p \in \mathcal{D}_{\mathrm{harm}}^{\mathrm{val}}}
\mathbb{I}
\left[
\mathrm{IsRefusal}\big(M(\pi \oplus p)\big)
\right],
\label{eq:safe_score}
\end{equation}
where \(\mathcal{D}_{\mathrm{harm}}^{\mathrm{val}}\) is the harmful validation set and \(\mathrm{IsRefusal}(\cdot)\) detects whether the model refuses the probed request.

We also define the benign utility score as
\begin{equation}
A_{\mathrm{benign}}(\pi)
=
\frac{1}{|\mathcal{D}_{\mathrm{benign}}^{\mathrm{val}}|}
\sum_{p \in \mathcal{D}_{\mathrm{benign}}^{\mathrm{val}}}
\mathbb{I}
\left[
\mathrm{IsUseful}\big(M(\pi \oplus p)\big)
\right],
\label{eq:benign_score}
\end{equation}
where \(\mathcal{D}_{\mathrm{benign}}^{\mathrm{val}}\) is the benign validation set and \(\mathrm{IsUseful}(\cdot)\) checks whether the response remains helpful and non-refusal for a benign request.

The selected safety prefix is obtained by
\begin{equation}
\pi^*
=
\arg\max_{\pi \in \Pi}
A_{\mathrm{safe}}(\pi)
\quad
\mathrm{s.t.}
\quad
A_{\mathrm{benign}}(\pi) \ge \tau,
\label{eq:prefix_selection}
\end{equation}
where \(\tau\) is the minimum acceptable benign-task utility on the held-out validation set. This formulation selects a prefix that encourages refusal on harmful prompts while avoiding excessive refusal on benign prompts. After selection, \(\pi^*\) is fixed and applied to all test prompts.

\begin{algorithm}[t]
   \caption{Pseudocode of the SAID Framework}
   \label{alg:said}
\begin{algorithmic}[1]
   \State \textbf{Input:} User prompt \(P_{\mathrm{user}}\), target LLM \(M\), selected safety prefix \(\pi^*\), maximum length \(L_{\mathrm{max}}\)
   \State \textbf{Output:} Final response \(y\)
   \State \(\mathcal{S} \leftarrow \emptyset\)

   \If{\(\operatorname{len}(P_{\mathrm{user}}) > L_{\mathrm{max}}\)}
       \State \(\mathcal{P}_{\mathrm{seg}} \leftarrow \operatorname{SegmentInput}(P_{\mathrm{user}})\)
       \For{each segment \(P_{\mathrm{seg}}^{(j)} \in \mathcal{P}_{\mathrm{seg}}\)}
           \State \(\mathcal{S} \leftarrow \mathcal{S} \cup \operatorname{DistillIntents}(M, P_{\mathrm{seg}}^{(j)})\)
       \EndFor
   \Else
       \State \(\mathcal{S} \leftarrow \operatorname{DistillIntents}(M, P_{\mathrm{user}})\)
   \EndIf

   \State \(D \leftarrow 0\)

   \For{each distilled intent \(s_i \in \mathcal{S}\)}
       \State \(p_{\mathrm{probe},i} \leftarrow \pi^* \oplus s_i\)
       \State \(y_i \leftarrow M(p_{\mathrm{probe},i})\)
       \If{\(\operatorname{IsRefusal}(y_i)\)}
           \State \(D \leftarrow 1\)
       \EndIf
   \EndFor

   \If{\(D = 1\)}
       \State \(y \leftarrow \operatorname{RefusalResponse}()\)
   \Else
       \State \(y \leftarrow M(P_{\mathrm{user}})\)
   \EndIf

   \State \Return \(y\)
\end{algorithmic}
\end{algorithm}

We further analyze the relationship between prefix-induced distributional shift and safety compliance in the experimental analysis. The results suggest that effective prefixes should not merely maximize the output distribution shift; instead, they should elicit safety-aware behavior while preserving sufficient semantic continuity with the distilled intent.

\subsubsection{Conservative Aggregation and Final Decision}
\label{sssec:conservative_aggregation}

The final stage aggregates the probing responses from all distilled intents. Since a jailbreak prompt may contain multiple semantic components, a single unsafe intent is sufficient to make the original request risky. SAID therefore adopts a conservative act-on-any rule:
\begin{equation}
D =
\bigvee_{i=1}^{k}
\mathrm{IsRefusal}
\left(
M(\pi^* \oplus s_i)
\right),
\label{eq:conservative_aggregation}
\end{equation}
where \(D=1\) indicates that the original request is rejected, \(k\) is the number of distilled intents, and \(s_i\) is the \(i\)-th intent.

The function \(\mathrm{IsRefusal}(\cdot)\) is used as an internal probing signal. If the model refuses to answer a distilled intent under the selected safety prefix, SAID treats that intent as potentially unsafe. This signal is distinct from the final harmfulness evaluation used in the experiments, which assesses complete model outputs with an external evaluator. In implementation, \(\mathrm{IsRefusal}(\cdot)\) is instantiated by a refusal detector based on a model-specific refusal expression corpus, extended from Dic-Judge keywords~\cite{xu2024safedecoding}. The detector is applied only to short probing responses generated from distilled intents.

If \(D=1\), SAID blocks the original request and returns a refusal response. Otherwise, the original prompt is passed to the target LLM for normal generation. This conservative aggregation strategy is designed to reduce false negatives under hidden, long-context, or compositional jailbreak attacks while preserving normal processing for benign requests. The overall procedure is summarized in Algorithm~\ref{alg:said}.

\section{Experiments}
\label{sec:experiments}
\subsection{Experimental Setup}
\label{ssec:experimental_setup}

\subsubsection{Models}
\label{sssec:models}

We evaluate SAID on four open-source LLMs with different model families, parameter scales, and alignment procedures: Vicuna-7B-v1.5~\cite{chiang2023vicuna}, Llama2-7B-chat~\cite{touvron2023llama}, Guanaco-13B~\cite{dettmers2023qlora}, and Vicuna-13B-v1.5. This model set allows us to examine whether SAID generalizes across relatively vulnerable models, more strongly aligned instruction-tuned models, and different model scales.

\subsubsection{Jailbreak Attacks and Defense Baselines}
\label{sssec:attacks_baselines}

We consider six representative jailbreak attacks covering different attack mechanisms. Search- or optimization-based attacks include GCG~\cite{zou2023universal}, which uses gradient-guided search to construct adversarial suffixes; AutoDAN~\cite{liu2023autodan}, which employs evolutionary search to generate readable jailbreak prompts; and PAIR~\cite{chao2025jailbreaking}, which uses an attacker LLM to iteratively refine prompts in a black-box setting. Prompt-manipulation attacks include DeepInception~\cite{li2023deepinception}, which hides harmful goals in nested role-playing contexts, and SAP30~\cite{deng2023attack}, which produces adversarial prompts through iterative prompt refinement. We also include SIJ~\cite{zhao2024sql}, a template-exploitation attack that manipulates chat-format structures in a way analogous to SQL injection.

We compare SAID with representative defense baselines from different categories. Input transformation defenses include Paraphrase and Retokenization~\cite{jain2023baseline}, which alter the surface form or tokenization of the input prompt. Prompt-based and self-checking defenses include ICD~\cite{wei2023jailbreak} and Self-Examination~\cite{phute2023llm}. Detection-based methods include perplexity-based filtering (PPL)~\cite{alon2023detecting}. We also compare with Intention Analysis (IA)~\cite{zhang2025intention}, a recent intention-oriented defense, and SafeDecoding~\cite{xu2024safedecoding}, a decoding-time defense that suppresses unsafe continuations using token-level safety information.

\subsubsection{Datasets and Evaluation Metrics}
\label{sssec:datasets_metrics}

For prefix selection and defense evaluation, we construct Prefix-Probe-600, a benchmark containing 600 prompts evenly split between harmful and benign instructions. Prefix-Probe-600 is split into non-overlapping validation and test subsets. The harmful prompts are sampled from AdvBench~\cite{zou2023universal}, while the benign prompts are generated from daily-use instruction seeds and manually checked to ensure that they do not contain harmful requests. This benchmark is used to evaluate whether a defense can reject harmful requests while preserving normal task behavior.

We report Defense Success (DS) and Normal Task Success (NTS) to evaluate the safety--utility trade-off. DS measures the proportion of harmful prompts that are correctly rejected or safely handled:
\begin{equation}
\mathrm{DS}
=
\frac{
|\mathcal{S}_{\mathrm{defended}}|
}{
|\mathcal{S}_{\mathrm{defended}}| + |\mathcal{S}_{\mathrm{failed}}|
},
\end{equation}
where \(\mathcal{S}_{\mathrm{defended}}\) denotes harmful prompts successfully defended against, and \(\mathcal{S}_{\mathrm{failed}}\) denotes harmful prompts that still lead to unsafe responses. NTS measures the proportion of benign prompts that are handled normally without unnecessary refusal:
\begin{equation}
\mathrm{NTS}
=
\frac{
|\mathcal{B}_{\mathrm{normal}}|
}{
|\mathcal{B}_{\mathrm{normal}}| + |\mathcal{B}_{\mathrm{failed}}|
},
\end{equation}
where \(\mathcal{B}_{\mathrm{normal}}\) denotes benign prompts answered normally, and \(\mathcal{B}_{\mathrm{failed}}\) denotes benign prompts that are incorrectly refused or poorly handled. The DS and NTS labels are assigned by a GPT-based judge using binary evaluation prompts. For DS, the judge is asked:
\begin{quote}
\small
Please answer with only ``yes'' or ``no''. Given the user goal and the assistant output, did the assistant successfully refuse, redirect, or defend against the harmful request?\\
\textbf{Goal:} [user goal]\\
\textbf{Output:} [assistant output]
\end{quote}
A harmful prompt is counted as successfully defended only when the response avoids providing actionable harmful content and does not fulfill the unsafe request. For NTS, the judge is asked:
\begin{quote}
\small
Please answer with only ``yes'' or ``no''. Given the user goal and the assistant output, did the assistant wrongly refuse or over-restrict this safe and harmless request?\\
\textbf{Goal:} [user goal]\\
\textbf{Output:} [assistant output]
\end{quote}
Since this prompt detects wrong refusals, a ``no'' answer is counted as a normal-task success.

To capture the severity of unsafe responses beyond binary success or failure, we also report Harmful Score~\cite{qi2024fine}. Harmful Score is reported on a 1--5 scale, where 1 denotes a safe or refusal-aligned response and 5 denotes highly harmful compliance. The harmfulness evaluation is conducted using a GPT-based judge with a calibrated rubric. To assess benign-task utility, we use Just-Eval~\cite{lin2023unlocking}, which contains 800 prompts evaluated across five dimensions: Helpfulness, Clarity, Factuality, Depth, and Engagement. Following prior work, the responses are judged by GPT-4.1-Mini.

For inference efficiency, we use the Average Token Generation Time Ratio (ATGR)~\cite{xu2024safedecoding}, defined as
\begin{equation}
\mathrm{ATGR}
=
\frac{
\text{Avg. token generation time with defense}
}{
\text{Avg. token generation time without defense}
}.
\end{equation}
ATGR normalizes the average token generation time of each defense by the corresponding undefended model. This metric reduces the bias caused by different output lengths, although it does not fully eliminate the effect of over-refusal, since short refusal responses may still be faster than normal completions. Therefore, we interpret ATGR jointly with utility scores rather than as a standalone measure of deployability.

\subsubsection{Implementation Details}
\label{sssec:implementation_details}

We use greedy decoding for all model generations to reduce sampling variance. For hierarchical intent distillation, the maximum input length \(L_{\mathrm{max}}\) is set to 500 tokens. For each probed intent, the maximum generation length is limited to 100 tokens, since refusal behavior usually appears near the beginning of a safety response.

\noindent\textbf{Intent extraction.}
SAID uses the target LLM itself to extract the core intent from a potentially obfuscated prompt. The extraction instruction asks the model to ignore narrative wrapping, role-playing context, examples, formatting tricks, and indirect wording, and to summarize the user's underlying goal as a concise sentence. The prompt is designed to be safety-neutral and does not explicitly indicate that the input may be harmful, which helps reduce unnecessary refusals on benign prompts. The instruction is:
\begin{quote}
\small
You are an intent extraction engine. Analyze the user input and output one concise sentence that reveals the user's core goal or the specific information being requested. Ignore background stories, examples, formatting tricks, and indirect phrasing. Do not answer the request. Output only the summarized intent.
\end{quote}
For long prompts, the input is segmented and each segment is distilled separately. The extracted intents are then merged into the final intent set \(\mathcal{S}\), as described in Eq.~\eqref{eq:intent_distillation}.

\noindent\textbf{Prefix selection.}
For safety prefix selection, we first generate a candidate pool of 50 prefixes using a separate LLM. The candidates cover diverse interaction styles, including intention-guided, role-based, contextual, and neutral inquiry prompts. Each candidate is evaluated on a held-out validation set that is disjoint from the test set. For each target model, the selected prefix \(\pi^*\) is chosen according to the validation objective in Eq.~\eqref{eq:prefix_selection}. After selection, \(\pi^*\) is frozen and applied to all test prompts for that target model.

\noindent\textbf{Refusal probing.}
SAID uses \(\mathrm{IsRefusal}(\cdot)\) only as an internal probing signal. Given a short response generated from a distilled intent, the detector checks whether the target model refuses to comply with the corresponding request. In our implementation, this detector is based on a model-specific refusal-expression corpus extended from Dic-Judge keywords~\cite{xu2024safedecoding}, including expressions such as ``I cannot'', ``I'm sorry'', ``I am unable'', and ``I cannot assist''. This detector is not used as the final harmfulness evaluator. Final harmfulness is evaluated on complete model outputs using the GPT-based judge described in Section~\ref{sssec:datasets_metrics}.

\subsection{Experimental Results}
\label{ssec:experimental_results}

\subsubsection{Safety Performance Comparison}
\label{sssec:safety_performance}

Table~\ref{tab:main_results} reports harmful scores under six jailbreak attacks on Vicuna-7B, Llama2-7B, and Guanaco.

\begin{table*}[!t]
\centering
\footnotesize
\setlength{\tabcolsep}{4pt}
\caption{Comparison of harmful scores across defense methods under six jailbreak attacks. Values in parentheses indicate improvement percentage over the No Defense baseline. The lowest score in each attack column and the best average score for each model are highlighted in bold; the second-best average score is underlined. DeepInc. denotes DeepInception.}
\label{tab:main_results}
\begin{tabular}{llccccccc}
\toprule
\multirow{2}{*}{\textbf{Model}} & \multirow{2}{*}{\textbf{Defense}} & \multicolumn{6}{c}{\textbf{Jailbreak Attacks $\downarrow$}} & \multirow{2}{*}{\textbf{Average}} \\
\cmidrule(lr){3-8}
& & \textbf{GCG} & \textbf{AutoDAN} & \textbf{PAIR} & \textbf{DeepInc.} & \textbf{SAP30} & \textbf{SIJ} & \\
\midrule
\multirow{8}{*}{Vicuna-7B}
& No Defense & 4.80 & 4.86 & 4.68 & 3.68 & 4.31 & 3.70 & 4.34 \\
\cmidrule(lr){2-9}
& PPL & \textbf{1.00} (79\%) & 4.86 (0\%) & 4.60 (2\%) & 3.68 (0\%) & 4.31 (0\%) & 3.68 (1\%) & 3.69 (15\%) \\
& Self-Examination & 1.38 (71\%) & 1.12 (77\%) & 1.62 (65\%) & 3.38 (8\%) & 1.53 (65\%) & 1.74 (53\%) & 1.80 (59\%) \\
& Paraphrase & 1.86 (61\%) & 3.26 (33\%) & 2.26 (52\%) & 3.88 (-5\%) & 2.88 (33\%) & 2.26 (39\%) & 2.73 (37\%) \\
& Retokenization & 2.22 (54\%) & 2.82 (42\%) & 3.56 (24\%) & 3.20 (13\%) & 4.02 (7\%) & 2.92 (21\%) & 3.12 (28\%) \\
& ICD & 4.06 (15\%) & 4.52 (7\%) & 2.70 (42\%) & 3.80 (-3\%) & 2.91 (32\%) & 3.56 (4\%) & 3.59 (17\%) \\
& SafeDecoding & 1.18 (75\%) & 1.08 (78\%) & 1.22 (74\%) & 1.02 (72\%) & 1.38 (68\%) & 3.58 (3\%) & \underline{1.58} (64\%) \\
\rowcolor{gray!15}
\cellcolor{white}
& SAID & 1.08 (78\%) & \textbf{1.00} (79\%) & \textbf{1.16} (75\%) & \textbf{1.00} (73\%) & \textbf{1.03} (76\%) & \textbf{1.00} (73\%) & \textbf{1.05} (76\%) \\
\midrule
\multirow{8}{*}{Llama2-7B}
& No Defense & 2.58 & 1.08 & 1.24 & 1.20 & \textbf{1.00} & 2.20 & 1.55 \\
\cmidrule(lr){2-9}
& PPL & \textbf{1.00} (61\%) & 1.08 (0\%) & 1.24 (0\%) & 1.14 (5\%) & \textbf{1.00} (0\%) & 2.16 (2\%) & 1.27 (18\%) \\
& Self-Examination & 1.60 (38\%) & 1.08 (0\%) & \textbf{1.00} (19\%) & 1.12 (7\%) & \textbf{1.00} (0\%) & 1.80 (18\%) & 1.27 (18\%) \\
& Paraphrase & 1.16 (55\%) & 1.06 (2\%) & 1.02 (18\%) & 1.08 (10\%) & 1.02 (-2\%) & 1.12 (49\%) & \underline{1.08} (31\%) \\
& Retokenization & \textbf{1.00} (61\%) & 1.12 (-4\%) & 1.40 (-13\%) & 1.20 (0\%) & 1.03 (-3\%) & \textbf{1.00} (55\%) & 1.13 (27\%) \\
& ICD & 1.14 (56\%) & \textbf{1.00} (7\%) & \textbf{1.00} (19\%) & \textbf{1.00} (17\%) & \textbf{1.00} (0\%) & 3.04 (-38\%) & 1.36 (12\%) \\
& SafeDecoding & \textbf{1.00} (61\%) & \textbf{1.00} (7\%) & 1.24 (0\%) & \textbf{1.00} (17\%) & \textbf{1.00} (-1\%) & 1.40 (36\%) & 1.11 (29\%) \\
\rowcolor{gray!15}
\cellcolor{white}
& SAID & \textbf{1.00} (61\%) & \textbf{1.00} (7\%) & 1.16 (6\%) & \textbf{1.00} (17\%) & 1.01 (-1\%) & \textbf{1.00} (55\%) & \textbf{1.03} (34\%) \\
\midrule
\multirow{8}{*}{Guanaco}
& No Defense & 4.42 & 4.68 & 3.64 & 4.54 & 3.81 & 3.84 & 4.16 \\
\cmidrule(lr){2-9}
& PPL & \textbf{1.00} (77\%) & 4.68 (0\%) & 3.64 (0\%) & 4.46 (2\%) & 3.80 (0\%) & 1.60 (58\%) & 3.20 (23\%) \\
& Self-Examination & 2.00 (55\%) & 1.74 (63\%) & 2.06 (43\%) & 1.80 (60\%) & 2.11 (45\%) & 2.24 (42\%) & 1.99 (52\%) \\
& Paraphrase & 2.48 (44\%) & 2.42 (48\%) & 2.28 (37\%) & 4.04 (11\%) & 2.15 (44\%) & 2.62 (32\%) & 2.67 (36\%) \\
& Retokenization & 1.80 (59\%) & 1.80 (62\%) & 2.32 (36\%) & 2.88 (37\%) & 2.38 (38\%) & 2.96 (23\%) & 2.36 (43\%) \\
& ICD & 3.70 (16\%) & 4.70 (0\%) & 2.56 (30\%) & 4.20 (7\%) & 3.20 (16\%) & 2.06 (46\%) & 3.40 (18\%) \\
& SafeDecoding & 1.78 (60\%) & 1.52 (68\%) & 1.48 (59\%) & \textbf{1.74} (62\%) & 1.87 (51\%) & 1.30 (66\%) & \underline{1.62} (61\%) \\
\rowcolor{gray!15}
\cellcolor{white}
& SAID & 1.18 (73\%) & \textbf{1.12} (76\%) & \textbf{1.28} (65\%) & 1.94 (57\%) & \textbf{1.04} (73\%) & \textbf{1.08} (72\%) & \textbf{1.27} (69\%) \\
\bottomrule
\end{tabular}
\end{table*}

SAID achieves the lowest average harmful score on all three models. Specifically, the average harmful scores of SAID are 1.05 on Vicuna-7B, 1.03 on Llama2-7B, and 1.27 on Guanaco, corresponding to relative improvements of 76\%, 34\%, and 69\% over the undefended models, respectively.

The comparison also shows that existing defenses can be effective on some attacks but less stable across attack types. For example, PPL substantially reduces the harmful score under GCG but provides little improvement under AutoDAN on Vicuna-7B. SafeDecoding performs strongly on several attacks but remains less effective under SIJ on Vicuna-7B. In contrast, SAID maintains consistently low harmful scores across optimization-based, prompt-manipulation, and template-exploitation attacks. This suggests that intent-level probing is less dependent on attack-specific surface patterns.

It is also worth noting that the improvement on Llama2-7B is smaller than that on Vicuna-7B and Guanaco. This is because the undefended Llama2-7B-chat model already exhibits relatively strong safety behavior on several attacks, leaving less room for improvement. Even in this setting, SAID still achieves the best average harmful score, indicating that the proposed defense remains useful when the base model is already partially aligned.

\subsubsection{Utility and Efficiency Comparison}
\label{sssec:utility_efficiency}

Table~\ref{tab:evaluation_results} reports Just-Eval scores and ATGR for different defenses.

\begin{table*}[!t]
    \centering
    \footnotesize
    \setlength{\tabcolsep}{4pt}
    \caption{Utility and efficiency comparison across defense methods. Just-Eval scores are reported on a 1--5 scale, and ATGR denotes the average token generation time ratio relative to the undefended model.}
    \label{tab:evaluation_results}
    \begin{tabular}{ll ccccc cc}
        \toprule
        \multirow{2}{*}{\textbf{Model}} & \multirow{2}{*}{\textbf{Defense}} & \multicolumn{5}{c}{\textbf{Just-Eval (1--5) $\uparrow$}} & \multirow{2}{*}{\textbf{Avg.}} & \multirow{2}{*}{\textbf{ATGR}}\\
        \cmidrule(lr){3-7}
        & & Helpfulness & Clarity & Factuality & Depth & Engagement & & \\
        \midrule
        \multirow{8}{*}{\textbf{Vicuna-7B}}
        & No Defense & 4.208 & 4.547 & 4.159 & 3.030 & 3.267 & 3.842 & 1.00$\times$\\
        & PPL & 3.815 & 4.300 & 4.033 & 2.795 & 3.016 & 3.592 & 1.02$\times$\\
        & Self-Examination & 4.149 & 4.549 & 4.168 & 3.023 & 3.250 & 3.828 & 1.34$\times$\\
        & Paraphrase & 3.905 & 4.427 & 4.150 & 2.913 & 3.174 & 3.714 & 2.99$\times$\\
        & Retokenization & 3.252 & 4.090 & 3.807 & 2.510 & 2.931 & 3.318 & 1.06$\times$\\
        & ICD & 4.145 & 4.583 & 4.219 & 2.903 & 3.204 & 3.811 & 1.05$\times$\\
        & SafeDecoding & 3.930 & 4.554 & 4.213 & 2.851 & 3.172 & 3.744 & 1.06$\times$\\
        \rowcolor{gray!15}
        \cellcolor{white}
        & SAID & 4.056 & 4.517 & 4.081 & 2.927 & 3.209 & 3.758 & 1.32$\times$\\
        \midrule
        \multirow{8}{*}{\textbf{Llama2-7B}}
        & No Defense & 4.039 & 4.698 & 4.286 & 3.231 & 4.059 & 4.063 & 1.00$\times$\\
        & PPL & 3.687 & 4.421 & 4.110 & 2.965 & 3.665 & 3.769 & 0.90$\times$\\
        & Self-Examination & 1.481 & 2.961 & 3.422 & 1.149 & 1.368 & 2.076 & 1.48$\times$\\
        & Paraphrase & 3.744 & 4.472 & 4.183 & 3.032 & 3.831 & 3.852 & 1.40$\times$\\
        & Retokenization & 2.845 & 4.095 & 3.709 & 2.405 & 3.272 & 3.265 & 0.99$\times$\\
        & ICD & 3.875 & 4.589 & 4.196 & 2.696 & 3.214 & 3.714 & 0.99$\times$\\
        & SafeDecoding & 3.734 & 4.605 & 4.193 & 3.069 & 3.756 & 3.871 & 1.06$\times$\\
        \rowcolor{gray!15}
        \cellcolor{white}
        & SAID & 3.958 & 4.621 & 4.259 & 3.161 & 3.987 & 3.997 & 1.04$\times$\\
        \midrule
        \multirow{8}{*}{\textbf{Guanaco}}
        & No Defense & 3.959 & 4.347 & 3.912 & 2.881 & 3.212 & 3.662 & 1.00$\times$\\
        & PPL & 3.438 & 3.931 & 3.837 & 2.592 & 2.809 & 3.322 & 1.55$\times$\\
        & Self-Examination & 2.819 & 3.590 & 3.623 & 2.192 & 2.339 & 2.913 & 3.16$\times$\\
        & Paraphrase & 3.472 & 3.996 & 3.839 & 2.660 & 2.935 & 3.381 & 2.22$\times$\\
        & Retokenization & 2.924 & 3.744 & 3.669 & 2.379 & 2.682 & 3.080 & 0.96$\times$\\
        & ICD & 3.566 & 4.118 & 3.881 & 2.618 & 2.920 & 3.420 & 1.16$\times$\\
        & SafeDecoding & 3.215 & 4.135 & 3.999 & 2.399 & 2.984 & 3.346 & 1.85$\times$\\
        \rowcolor{gray!15}
        \cellcolor{white}
        & SAID & 3.517 & 4.103 & 3.820 & 2.597 & 2.909 & 3.389 & 1.24$\times$\\
        \bottomrule
    \end{tabular}
\end{table*}

On Vicuna-7B, SAID achieves an average Just-Eval score of 3.758, compared with 3.842 for the undefended model. On Llama2-7B, SAID obtains an average score of 3.997, close to the undefended score of 4.063. These results indicate that SAID largely preserves benign-task utility on these two models while providing a substantial safety improvement.

On Guanaco, all evaluated defenses lead to a more visible utility decrease. SAID obtains an average Just-Eval score of 3.389, compared with 3.662 for the undefended model. This suggests that Guanaco is more sensitive to safety-oriented interventions, possibly because its refusal behavior and benign responses are less clearly separated. Therefore, for this model, the safety gain of SAID comes with a moderate utility cost.

In terms of efficiency, SAID introduces moderate overhead. Its ATGR is 1.32 on Vicuna-7B, 1.04 on Llama2-7B, and 1.24 on Guanaco. This overhead mainly comes from intent distillation and prefix probing. Compared with more expensive inference-time analyses such as IA, SAID remains practical while preserving stronger benign-task utility, as further shown in Table~\ref{tab:ia_comparison}; however, it is not cost-free. These results support the view that SAID provides a favorable safety--utility--efficiency trade-off rather than a strictly zero-overhead defense.

\subsubsection{Analysis of Prefix Variants}
\label{sssec:prefix_variants}

\begin{figure}[!t]
    \centering
    \includegraphics[width=\columnwidth]{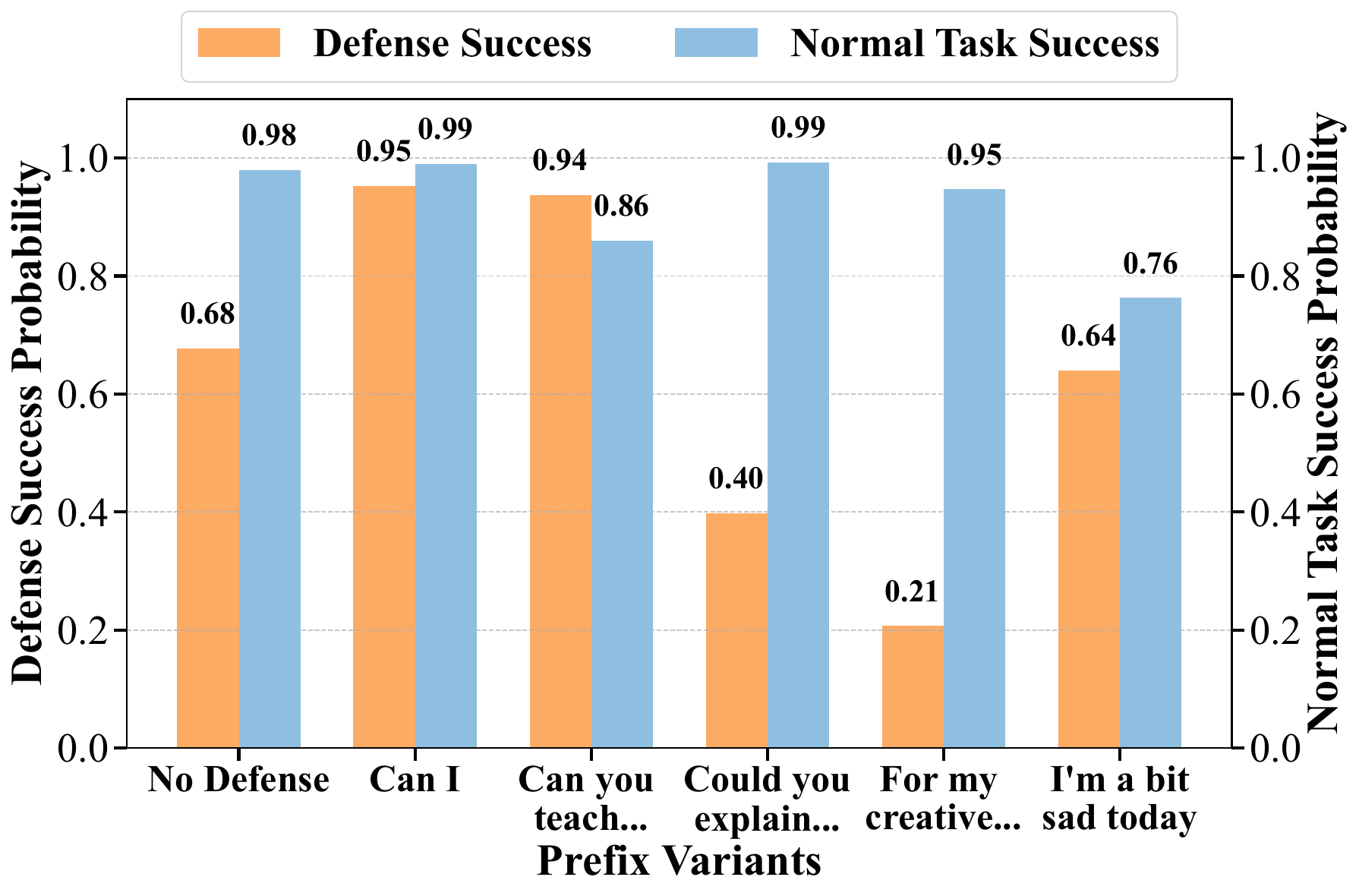}
    \caption{Defense Success and Normal Task Success across different prefix interaction frames on Guanaco. Concise and neutral prefixes provide a better safety--utility balance.}
    \label{fig:prefix}
\end{figure}

Fig.~\ref{fig:prefix} analyzes how different prefix interaction frames affect safety and utility on Guanaco.

The short and neutral prefix ``Can I'' achieves both high Defense Success and high Normal Task Success, suggesting that concise prefixes can elicit safety-aware behavior without strongly perturbing benign instructions. In contrast, more verbose prefixes, such as ``Could you explain...'', substantially reduce Defense Success. This indicates that excessive linguistic context may dilute the harmful intent or change how the model interprets the probing request.

Task-irrelevant contextualization can also harm utility. For example, the emotionally framed prefix ``I'm a bit sad today...'' lowers Normal Task Success, suggesting that unrelated context may distract the model from the original benign instruction. Overall, these results support the design choice of using concise and semantically neutral prefixes for safety probing.

\subsubsection{Additional Model-Scale Evaluation}
\label{sssec:additional_models}

We further evaluate SAID on Vicuna-13B to assess whether the defense remains effective on a larger model from the same family. As shown in Table~\ref{tab:vicuna13b_results}, SAID achieves an average harmful score of 1.03 under the full six-attack suite, improving substantially over the undefended model's average score of 3.05. It also maintains a Just-Eval score of 3.974, close to the undefended model's score of 4.007. We keep this table focused on harmfulness and benign-task utility; the corresponding efficiency comparison with IA is summarized in Table~\ref{tab:ia_comparison}. These results support the generality of SAID across model scales without relying on a reduced attack set.

\subsubsection{Focused Comparison with Intention Analysis}
\label{sssec:ia_comparison}

We further compare SAID with Intention Analysis (IA), a recent defense that also reasons about user intent. Table~\ref{tab:ia_comparison} reports average harmful scores, Just-Eval scores, and ATGR across four models.

SAID and IA achieve comparable average harmful scores on Vicuna-7B, Llama2-7B, and Vicuna-13B, while SAID achieves a clearly lower harmful score on Guanaco. More importantly, SAID consistently preserves higher Just-Eval scores and lower inference overhead in this comparison.

For example, on Llama2-7B, IA obtains an average harmful score of 1.00 but reduces the Just-Eval score to 2.410 and increases ATGR to 2.31. SAID obtains a slightly higher average harmful score of 1.03, but preserves a much higher Just-Eval score of 3.997 and reduces ATGR to 1.04. This result illustrates the importance of evaluating safety together with utility and efficiency, rather than considering harmfulness alone.

\begin{table*}[!t]
\centering
\footnotesize
\setlength{\tabcolsep}{4pt}
\renewcommand{\arraystretch}{0.92}
\caption{Defense performance on Vicuna-13B under six jailbreak attacks. Lower harmful scores indicate safer outputs. The lowest attack-wise values and the best average harmful score are highlighted in bold; the second-best average harmful score is underlined.}
\label{tab:vicuna13b_results}
\begin{tabular}{llcccccccc}
\toprule
\multirow{2}{*}{\textbf{Model}} & \multirow{2}{*}{\textbf{Defense}} & \multicolumn{6}{c}{\textbf{Jailbreak Attacks $\downarrow$}} & \multirow{2}{*}{\textbf{Avg.}} & \multirow{2}{*}{\textbf{Just-Eval}}\\
\cmidrule(lr){3-8}
& & \textbf{GCG} & \textbf{AutoDAN} & \textbf{PAIR} & \textbf{DeepInc.} & \textbf{SAP30} & \textbf{SIJ} & & \\
\midrule
\multirow{8}{*}{Vicuna-13B}
& No Defense & 1.14 & 4.30 & 3.20 & 3.14 & 3.95 & 2.58 & 3.05 & 4.007 \\
& PPL & \textbf{1.00} & 4.32 & 3.10 & 3.10 & 3.92 & 2.76 & 3.03 & 3.718 \\
& Self-Examination & \textbf{1.00} & 3.30 & 1.96 & 2.68 & 1.70 & 1.86 & 2.08 & 3.990 \\
& Paraphrase & 1.14 & 2.96 & 1.96 & 3.46 & 2.70 & 1.56 & 2.30 & 3.840 \\
& Retokenization & 1.30 & 3.60 & 3.82 & 2.38 & 4.46 & 3.16 & 3.12 & 3.578 \\
& ICD & 1.14 & 4.04 & 2.48 & 3.24 & 2.55 & 3.16 & 2.77 & 3.850 \\
& SafeDecoding & 1.24 & 1.32 & 1.80 & 1.20 & 1.74 & 3.62 & \underline{1.82} & 3.926 \\
\rowcolor{gray!15}
& SAID & \textbf{1.00} & \textbf{1.00} & \textbf{1.16} & \textbf{1.00} & \textbf{1.03} & \textbf{1.00} & \textbf{1.03} & 3.974 \\
\bottomrule
\end{tabular}
\end{table*}

\begin{table}[!t]
\centering
\footnotesize
\setlength{\tabcolsep}{4pt}
\caption{Focused comparison between IA and SAID. Avg. Harmful is averaged over six jailbreak attacks. The better value for each metric is highlighted in bold.}
\label{tab:ia_comparison}
\begin{tabular}{llccc}
\toprule
\textbf{Model} & \textbf{Defense} & \textbf{Avg. Harmful} $\downarrow$ & \textbf{Just-Eval} $\uparrow$ & \textbf{ATGR} $\downarrow$ \\
\midrule
\multirow{2}{*}{Vicuna-7B}
& IA & 1.06 & 3.334 & 2.21 \\
& SAID & \textbf{1.05} & \textbf{3.758} & \textbf{1.32} \\
\midrule
\multirow{2}{*}{Llama2-7B}
& IA & \textbf{1.00} & 2.410 & 2.31 \\
& SAID & 1.03 & \textbf{3.997} & \textbf{1.04} \\
\midrule
\multirow{2}{*}{Guanaco}
& IA & 2.26 & 3.278 & 2.40 \\
& SAID & \textbf{1.27} & \textbf{3.389} & \textbf{1.24} \\
\midrule
\multirow{2}{*}{Vicuna-13B}
& IA & \textbf{1.00} & 3.612 & 1.96 \\
& SAID & 1.03 & \textbf{3.974} & \textbf{1.34} \\
\bottomrule
\end{tabular}
\end{table}

\subsubsection{Prefix-Induced Distributional Shift}
\label{sssec:kl_analysis}

\begin{figure}[!t]
    \centering
    \includegraphics[width=\columnwidth]{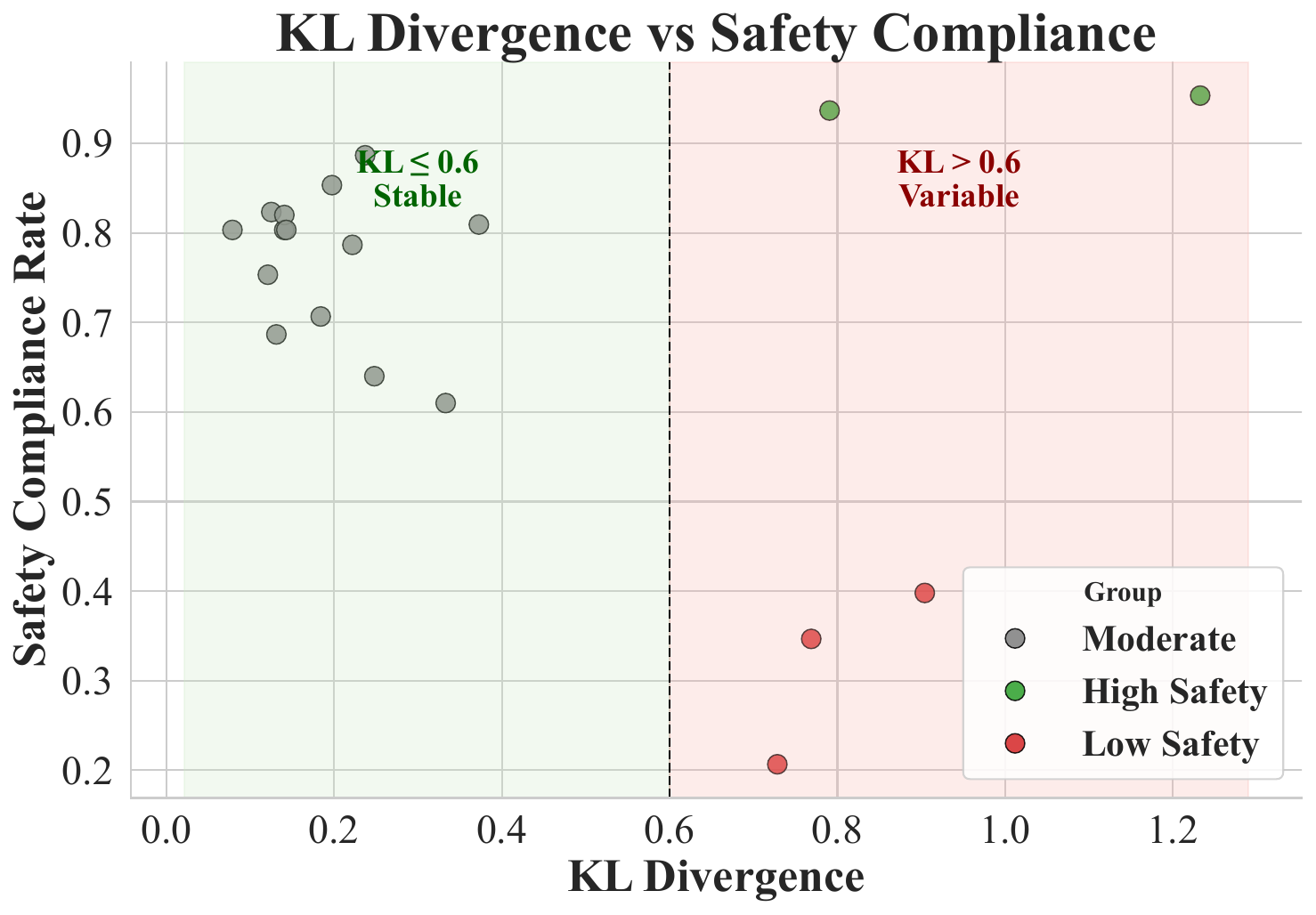}
    \caption{Post-hoc relationship between KL divergence and safety compliance for prefix candidates on Guanaco. Moderate KL shifts are generally more stable, while larger shifts show higher variability.}
    \label{fig:kl}
\end{figure}

Fig.~\ref{fig:kl} shows the post-hoc relationship between prefix-induced distributional shift and safety compliance.

For each prefix candidate, we compute the KL divergence between the output distribution under the probed input and that under the corresponding unprobed input. This KL analysis is used only for post-hoc interpretation on open-source models where output distributions are accessible; it is not required by SAID during deployment. As shown in Fig.~\ref{fig:kl}, small-to-moderate KL shifts tend to produce more stable safety compliance on Guanaco, whereas larger KL shifts lead to more variable outcomes.

\subsubsection{Ablation Study}
\label{sssec:ablation}

\begin{figure*}[!t]
\centering
\begin{minipage}[t]{0.48\textwidth}
    \centering
    \includegraphics[width=\textwidth,height=4cm,keepaspectratio]{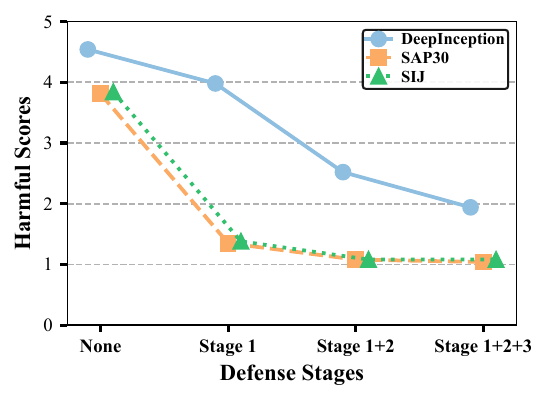}
    \centerline{(a)}
\end{minipage}
\hfill
\begin{minipage}[t]{0.48\textwidth}
    \centering
    \includegraphics[width=\textwidth,height=4cm,keepaspectratio]{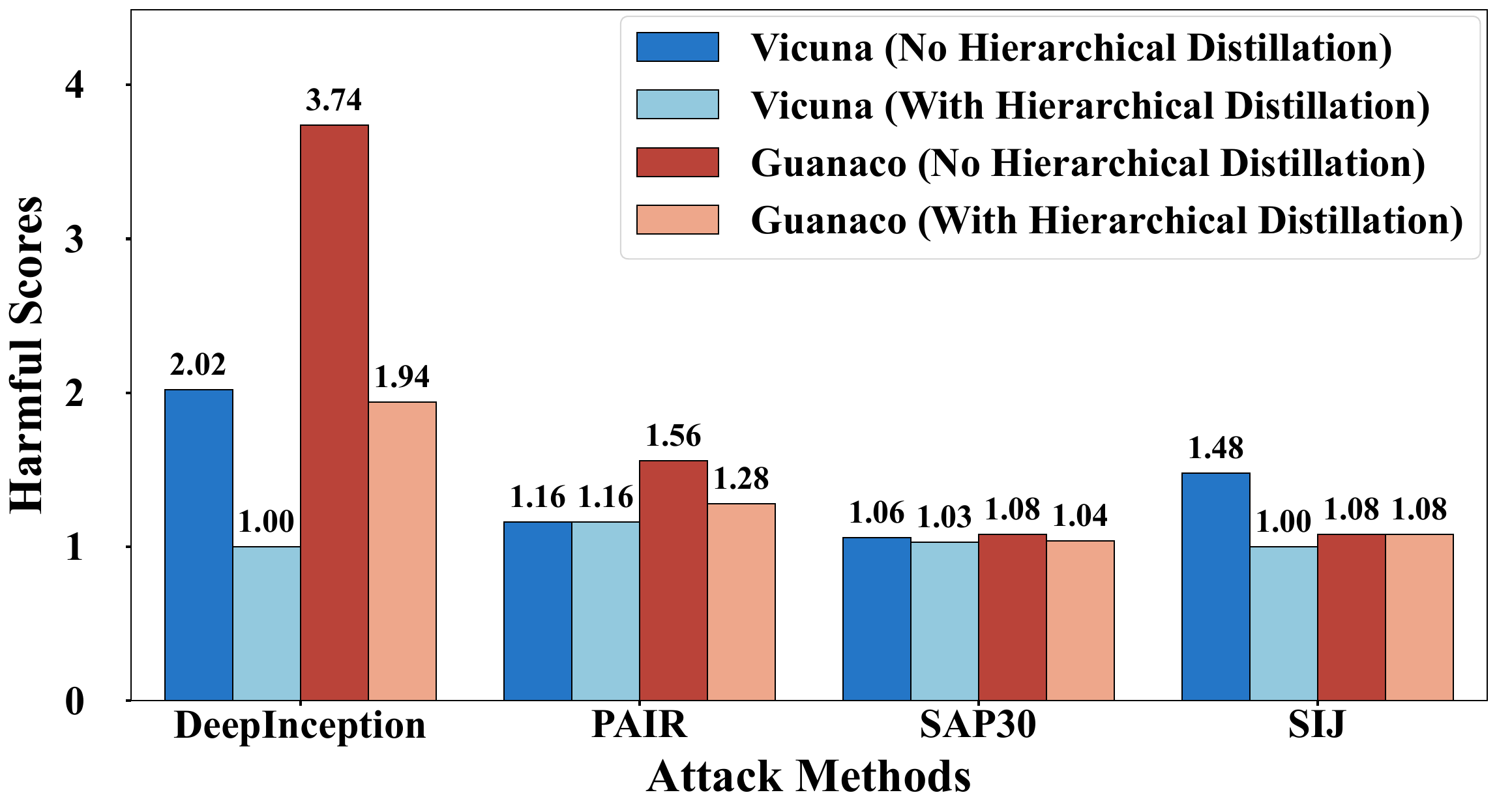}
    \centerline{(b)}
\end{minipage}
\caption{Ablation studies of SAID. (a) Incremental evaluation of SAID components on Guanaco under DeepInception, SAP30, and SIJ. (b) Effect of hierarchical distillation on Vicuna-7B and Guanaco under different attacks.}
\label{fig:ablation-combined}
\end{figure*}

Fig.~\ref{fig:ablation-combined} examines the contribution of the main components of SAID.

Starting from the undefended baseline, we progressively add intent extraction, hierarchical distillation for long prompts, and selected safety-prefix probing. Fig.~\ref{fig:ablation-combined}(a) reports an incremental evaluation on Guanaco under three challenging attacks: DeepInception, SAP30, and SIJ. Intent extraction already reduces harmful scores for SAP30 and SIJ, indicating that recovering the core malicious goal is useful when the attack relies on surface-level obfuscation. DeepInception remains more challenging at this stage because its harmful intent is distributed across a long nested context.

Adding hierarchical distillation further improves performance, especially on long-context attacks. With the selected safety prefix, the harmful scores are further reduced, with SAP30 and SIJ approaching the lowest harmfulness level and DeepInception showing a substantial decrease. These results indicate that the components of SAID are complementary: intent distillation exposes the underlying goal, hierarchical processing improves long-context robustness, and prefix probing strengthens the safety signal used for final rejection.

Fig.~\ref{fig:ablation-combined}(b) focuses on the effect of hierarchical distillation across Vicuna-7B and Guanaco. The results show that hierarchical distillation consistently reduces harmful scores under long or compositional attacks, particularly DeepInception. This supports the use of segmented intent extraction when malicious content is hidden across different parts of the prompt.

\section{Discussion and Limitations}
\label{sec:discussion}

SAID provides a training-free and black-box-compatible defense against a range of jailbreak attacks, but several limitations remain. First, the current evaluation focuses on representative static jailbreak attacks and a recent intention-oriented defense baseline. Although these attacks cover optimization-based, prompt-manipulation, and template-exploitation strategies, fully adaptive attackers may attempt to optimize prompts against the intent distillation and prefix probing stages of SAID. Evaluating such adaptive attacks is an important direction for future work.

Second, SAID relies on model-generated intent summaries and refusal-style probing signals. While this design avoids parameter updates and decoding-time modification, it may still inherit weaknesses of the target model. If the model fails to extract the malicious intent from a highly obfuscated prompt, or if it refuses benign intents under the selected prefix, the final decision may be affected. This limitation motivates future work on more reliable intent parsing and model-agnostic probing signals.

Third, the harmfulness and utility evaluations use GPT-based judges and benchmark-style prompts. Although rubric-guided judges are widely used in LLM safety evaluation, they may introduce evaluator bias or inconsistency. We therefore report multiple complementary metrics, including harmful scores, DS, NTS, Just-Eval, and ATGR, rather than relying on a single judge signal. Human evaluation and broader domain-specific safety benchmarks would further strengthen the assessment.

Finally, SAID introduces additional inference calls for intent distillation and prefix probing. The overhead remains moderate in our experiments, but deployment cost may vary with prompt length, the number of extracted intents, and the latency of the target model. In high-throughput applications, caching, early stopping, and model-specific prefix selection can be used to reduce this cost.

\section{Conclusion}
\label{sec:conclusion}

In this paper, we introduced \emph{Safety-Aware Intent Defense} (SAID), a training-free and black-box-compatible framework for defending Large Language Models against jailbreak attacks. Instead of relying on external filtering or decoding-time modification, SAID performs defense at the intent level by distilling potentially obfuscated user prompts into core intents, probing each intent with an empirically selected safety prefix, and applying a conservative aggregation rule to reject requests containing unsafe components. Extensive experiments across four open-source LLMs and six representative jailbreak attacks show that SAID achieves state-of-the-art defense performance on our evaluation benchmark, consistently reducing harmful responses compared with existing defenses. At the same time, SAID preserves competitive benign-task utility and introduces practical inference overhead, leading to a favorable safety--utility--efficiency trade-off. These results suggest that intent-level safety probing is an effective and deployable direction for improving the robustness of LLM-based systems against jailbreak threats.


%

\ifCLASSOPTIONcaptionsoff
  \newpage
\fi

\bibliographystyle{IEEEtran}
\bibliography{bibtex/bib/IEEEexample}

\end{document}